\input{aipcheck}

\def\ro{{\it ROSAT\/}}
\def\asca{{\it ASCA\/}}
\def\rxte{{\it RXTE\/}}

\def\cha{{\it Chandra\/}}

\documentclass[
    ,final            
  ]
  {aipproc}

\layoutstyle{6x9}

\begin{document}

\title{Was the X-ray Afterglow of GRB 970815 Detected?}

\author{N.\ Mirabal}{
  address={Astronomy Department, Columbia University, 550 West 120th Street,
 New York, NY 10027}
}

\author{
J. P.\ Halpern}{
  address={Astronomy Department, Columbia University, 550 West 120th Street,
 New York, NY 10027}
}

\author{E. V.\ Gotthelf}{
  address={Astronomy Department, Columbia University, 550 West 120th Street,
 New York, NY 10027}}

\iftrue
\author{R. Mukherjee}{
  address={Dept. of Physics \& Astronomy, Barnard College,
New York, NY 10027},
}

\begin{abstract}
GRB 970815 was a well-localized 
gamma-ray burst (GRB) detected by the All-Sky
Monitor (ASM) on the Rossi X-Ray Timing Explorer (RXTE) for which no 
afterglow was identified despite follow-up \asca\ and \ro\ pointings 
and optical imaging to limiting magnitude $R > 23$. While 
an X-ray source, AX/RX~J1606.8+8130, 
was detected just outside the ASM error box, it was never
associated with the GRB because it was not clearly fading and because no 
optical afterglow was ever discovered. We recently made 
deep optical observations of the
AX/RX~J1606.8+8130 position, which 
is blank to a limit of $V >$ 24.3 and $I > 24.0$, implying an
X-ray--to--optical flux ratio $f_{X}$/$f_{V}$ $>$ 500.
In view of this extreme limit, we analyze 
and reevaluate the \asca\ and \ro\ data and conclude 
that the X-ray source AX/RX~J1606.8+8130 was indeed the afterglow of
GRB 970815, which 
corresponds to an optically ``dark'' GRB. Alternatively, if AX/RX 
J1608+8130 is discovered to be a persistent
source, then it could be associated with EGRET source 3EG J1621+8203,
whose error box includes this position.
\end{abstract}

\date{\today}

\maketitle


\section{Introduction}

One of the most intriguing results from six years of GRB
follow-ups at optical wavelengths is that roughly 60$\%$ of 
well-localized GRBs lack an optical transient despite intensive 
ground-based searches (e.g. \cite{Fynbo:2001}).
Some of these ``dark'' GRBs could simply be due to a 
failure to image deeply or quickly enough. However, in certain
cases the afterglow may have been missed in the optical either because
it is obscured by dust in the host galaxy,
or because it is located at high-redshift ($z>5$).
We discuss here X-ray and optical observations of GRB 970815, which 
support an interpretation consistent with quite possibly the first
detection of a ``dark'' GRB in the afterglow era, preceding
GRB 970828 in that category \cite{Djo:2001}. 

\section{X-ray Observations}

GRB 970815 was localized by the ASM aboard RXTE on UT 1997
Aug. 15.50623, with a duration 
of $\approx$ 130 s \cite{Smith:1999}.
Simultaneous detection with two of the ASM scanning cameras 
 refined the position of GRB 970815 to
the small error box shown in Figure 1.
The superposed annulus based 
on the BATSE and Ulysses triangulation 
confirms the ASM position. Following the prompt localization by \rxte, two 
X-ray observations were made that cover the entire
\rxte\ error box, 
one by \asca\ \cite{Mur:1997} 
and one by the \ro\ High Resolution Imager (HRI) \cite{Greiner:1997}. 
Analysis of the data revealed no source brighter than
1 $\times 10^{-13}$~erg~cm$^{-2}$~s$^{-1}$ within the \rxte\ error box.  
There was, however, a 
source  AX/RX~J1606.8+8130 just outside the \rxte\
error box with an average flux 
F$_{x}$(2--10~keV) = $4.2 \times 10^{-13}$~erg~cm$^{-2}$~s$^{-1}$. 
Figure 1 shows the combined \asca\ SIS image and the location of
AX/RX~J1606.8+8130 with respect to the burst error box.
While AX/RX~J1606.8+8130 lies just outside the \rxte\ error box, it
is within the BATSE/Ulysses annulus. 
Another marginally significant  
\ro\ source RX~J1608.8+8131 lies inside the \rxte\ error box, but 
it was not detected in the earlier \asca\ observation.
Hereafter we concentrate our discussion on AX/RX~J1606.8+8130.

\begin{figure}
\includegraphics[height=.35\textheight,angle=-90]{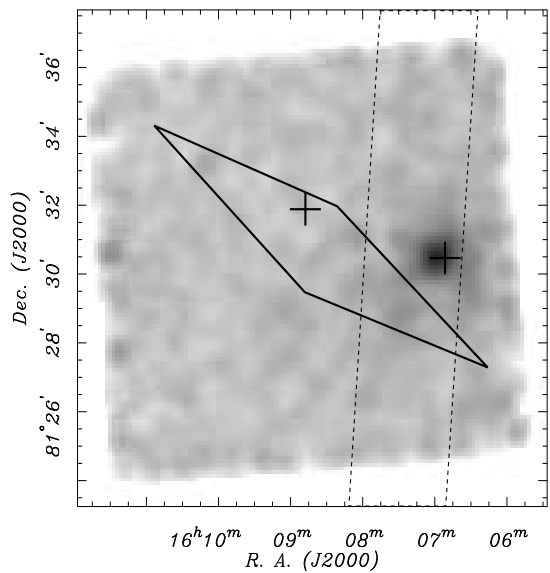}
  \caption{\asca\ CCD SIS image of the
field of GRB 970815, with the
\rxte\ ASM error box ({\it solid line})
and Ulysses/BATSE annulus ({\it dashed lines}) superposed.
\ro\ HRI point sources are indicated by {\it crosses}.}
\end{figure}

\section{X-ray Light Curve}


\begin{figure}
\includegraphics[height=.34\textheight]{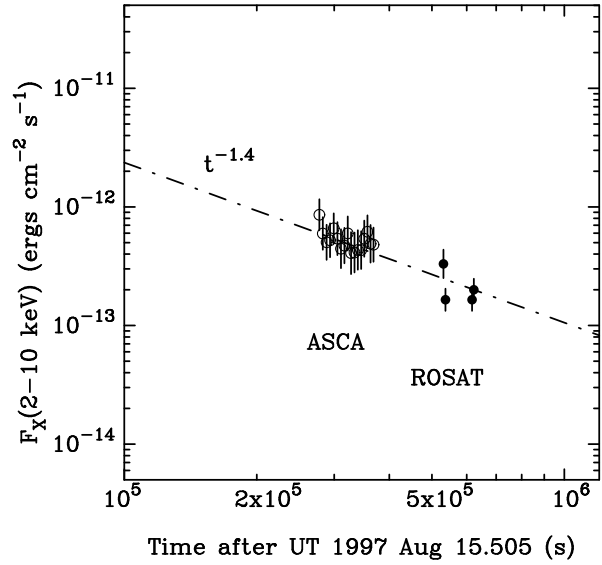}
  \caption{The X-ray light curve of AX/RX J1606.8+8130,
derived by assuming the simultaneous power-law fit to
the \asca\ spectra.
The {\it dash-dot\/} line shows a simple power-law
decay F$_{x}\propto$ t$^{-1.4}$, although the
variation in the \asca\ points are also consistent
with no overall decay.}
\end{figure}


Figure 2 shows the combined X-ray light curve of AX/RX~J1606.8+8130. 
The \asca~light curve includes
the sum of counts from all four detectors. The \ro\ points 
correspond to an extrapolated flux in the 2--10 keV band 
assuming the power-law spectral parameters derived from the 
\asca\ spectra ($\Gamma$ = 1.64 $\pm$ 0.35 and 
$N_{\rm H} <$ 1.3 $\times 10^{21}$~cm$^{-2}$), which might not be entirely 
valid if an
additional spectral component contributes significantly 
in the HRI soft band.  The individual \asca~and 
\ro~components of the light curve
show no obvious evidence for variability. However,
AX/RX~J1606.8+8130 is consistent with a 
F$_{x} \propto$ t$^{-1.4}$ flux decay between the \asca\ and \ro\ 
observations, easily within the range of well-studied 
GRB X-ray afterglows. Moreover, the integrated
2--10 keV  X-ray fluence corresponds to $\approx$ 10$\%$ of the GRB 
fluence, in agreement with the properties of other GRBs \cite{Front:2000}.

\section{Optical Observations of AX/RX~J1606.8+8130}

Following the rapid dissemination of the \rxte\ position for GRB 970815, 
a number of groups conducted 
optical imaging of its error box including the position of AX/RX~J1606.8+8130 
as early as 17 hr after the burst \cite{Harr:1997}. At the time, no 
significant variable sources were 
found at the X-ray position or within the \rxte\ error box 
to an upper limit $R > 23$ \cite{Harr:1997}. Years later while conducting a 
search for the $\gamma$-ray source 3EG J1621+8203 \cite{Muk:2002},
we reexamined the X-ray position
of AX/RX~J1606.8+8130 in several optical filters.
Figure 3 shows the adopted $10^{\prime\prime}$ radius \ro\ error
circle around the X-ray position, which is still optically blank to a 
3$\sigma$ limit of $V > 24.3$. 
In other filters, AX/RX J1606.8+8130 shows no evidence of a host galaxy or
any other optical counterpart to
limits of $B > 21.5$, $R > 22.0$, and $I > 24.0$.

\begin{figure}
\includegraphics[height=.3\textheight]{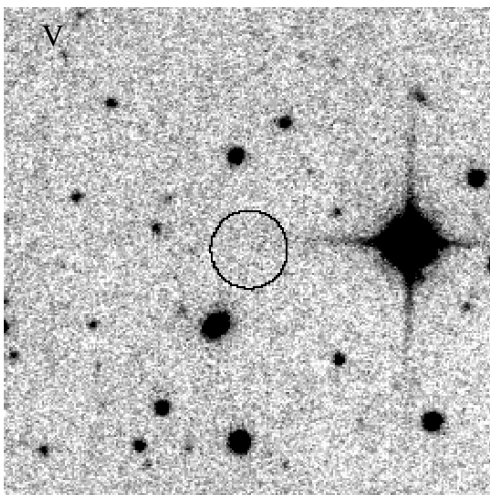}
  \caption{
Zoom-in on a deep $V$ image from the
MDM Observatory at the location of the unidentified
X-ray source AX/RX J1606.8+8130.
The field is $2^{\prime}$ across,
and the \ro\ HRI error circle is
$10^{\prime\prime}$ in radius.
The 3$\sigma$ upper limit is $V > 24.3$. North is up and east is to the left.}
\end{figure}

\section{Was AX/RX~J1606.8+8130 the afterglow of GRB 970815?}

Starting with  
the observed X-ray flux density $f_{X}$,  we can extrapolate a broad-band
spectrum of the form $f_{R}
= f_{X}(\nu_{R}/\nu_X)^{-\beta}$ where $f_{R}$ is the $R$-band 
optical flux density at a frequency $\nu_{R}$ and $\beta$ is the X-ray
spectral index. 
From the \asca\ spectra we have $f_{X} \approx 0.10\,\mu$Jy 
($\nu_{X}=4.84 \times 10^{17}$ Hz) at a time 
$t \sim 3.74$ days after the burst, and $\beta \approx
0.64$.  The optical flux density evolution would then correspond to
$f_{R}(t_{d}) \approx 55\,t_{d}^{-1.4}\,\mu$Jy
where $t_{d}$ is days
elapsed since the BATSE detection of GRB 970815. This translates into 
$R \approx 19.0$ on UT 1997 Aug. 16.31. 
Therefore, the predicted magnitude is brighter than the
$R > 23$ upper limit reported at that time \cite{Harr:1997}.
Such difference would tentatively 
support a ``dark'' GRB classification. Nonetheless,  
the chance superposition of 3EG J1621+8203 and  GRB 970815
introduces a slight doubt about the nature of AX/RX~J1606.8+8130.  
This is because ``dark'' GRBs can temporarily mimic the characteristics of 
a plausible counterpart for unidentified EGRET sources, namely, 
rotation-powered pulsars \cite{Mirabal:2001}.

\section{Conclusions and Future Work}

In summary, the properties of AX/RX~J1606.8+8130 support the idea 
that GRB 970815 corresponds to an optically ``dark'' GRB, quite possibly the
first detection of a ``dark'' GRB in the afterglow era, 
preceding
GRB 970828 in that category \cite{Djo:2001}. However, because of the 
chance superposition between 3EG J1621+8203 and GRB 970815, 
a slight doubt remains about the nature of AX/RX~J1606.8+8130.
\cha\ observations are planned that should resolve
the ambiguity of the possible connection between
AX/RX~J1606.8+8130 and either GRB 970815 or 3EG J1621+8203. 


\IfFileExists{\jobname.bbl}{}
 {\typeout{}
  \typeout{******************************************}
  \typeout{** Please run "bibtex \jobname" to optain}
  \typeout{** the bibliography and then re-run LaTeX}
  \typeout{** twice to fix the references!}
  \typeout{******************************************}
  \typeout{}
 }

\end{document}

\endinput